\documentclass[twocolumn]{aastex631}

\newcommand{\nircam}{NIRCam}

\newcommand{\sextractor}{\texttt{SExtractor}}

\definecolor{orange}{rgb}{0.89, 0.36, 0.1} 
\definecolor{pinky}{rgb}{0.8,0.04,0.8} 

\hypersetup{linkcolor=orange,citecolor=orange,filecolor=red,urlcolor=red}

\begin{document}

\title{A new era of intracluster light studies with JWST}

\correspondingauthor{Mireia Montes}
\email{mireia.montes.quiles@gmail.com}

\author[0000-0001-7847-0393]{Mireia Montes}

\affiliation{Instituto de Astrof\'{\i}sica de Canarias, c/ V\'{\i}a L\'actea s/n, E-38205 - La Laguna, Tenerife, Spain}
\affiliation{Departamento de Astrof\'isica, Universidad de La Laguna, E-38205 - La Laguna, Tenerife, Spain}   

\author[0000-0001-8647-2874]{Ignacio Trujillo}
\affiliation{Instituto de Astrof\'{\i}sica de Canarias, c/ V\'{\i}a L\'actea s/n, E-38205 - La Laguna, Tenerife, Spain}
\affiliation{Departamento de Astrof\'isica, Universidad de La Laguna, E-38205 - La Laguna, Tenerife, Spain}   



\begin{abstract}
Still largely unexplored, the diffuse light in cluster of galaxies traces the past and on-going buildup of these massive structures. Here, we present the first comprehensive study of the intracluster light (ICL) of the cluster SMACS-J0723.3-7327 (z=0.39) using the JWST Early Release Observations. These deep and high spatial resolution images allow the study of the ICL with high signal-to-noise ratio up to a radial distance of $\sim400$ kpc, twice as far with respect to previous HST studies of intermediate redshift clusters. This opens up the possibility of exploring the rich mixture of processes that are building the ICL. We find that the inner parts (R$<$100 kpc) are built through a major merger while the outer parts (R$>$ 100 kpc) are mainly produced by the tidal stripping of Milky Way-like satellites. We also found that the slope of the stellar mass density radial profile of the ICL of this cluster ($\alpha_{3D} = -2.47\pm0.13$) follows closely the predicted dark matter halo slope ($\alpha_{3D\mathrm{,DM}} = -2.6$ to $-2$), supporting the idea that both components have a similar shape and thus the potential of using the ICL as a tracer of the dark matter distribution in clusters of galaxies. Future JWST studies of the ICL are set to revolutionise our understanding of cluster formation and will be crucial to improve the gravitational lensing mass maps of these structures and thus to accurately characterise the properties of the first galaxies.
\end{abstract}

%


\section{Introduction} \label{sec:intro}

During the past 20 years, observations have shown that the ICL is a ubiquitous feature in galaxy clusters \citep{Feldmeier2002, Krick2007, Kluge2020}. During encounters between galaxies in groups and clusters, stars are freed from their hosts and end up populating the space between the galaxies. After some time, these unbound stars form the characteristic diffuse glow of the ICL. Consequently, this light is a signature of the assembly of clusters of galaxies \citep[see][for recent reviews]{Montes2019, Contini2021rev, Montes2022}.

The ICL has been shown to be an unexpected tool to study groups and clusters of galaxies in detail, allowing us to infer not only their assembly history but their radius and even dark matter distributions \citep[][]{MT19, Alonso-Asensio2020, Deason2021, Gonzalez2021}. For example, \citet{Mahler2022} showed that the inclusion of the ICL as a prior to model the mass distribution of the cluster SMACS-J0723.3-7327 resulted in more accurate mass maps (1\farcs08 of r.m.s compared to the 1\farcs26 of r.m.s without ICL). These results show the extraordinary potential of using the ICL as an accurate dark matter tracer. 

Observations of the ICL of intermediate redshift clusters have been, for the most part, limited to the blue part of the spectrum. These observations show clear radial gradients in colors indicating radial gradients in metallicity \citep[e.g.,][]{Mihos2017, Iodice2017, DeMaio2018, Montes2021} and, in some cases, age \citep[e.g.,][]{MT14, MT18, Morishita2017}. Consequently, exploring the ICL distribution and mass fraction of intermediate redshift clusters using only blue filters is prone to artificial biases. In this context, it is clear that deep infrared (IR) imaging with high spatial resolution, as the one provided by JWST \citep{Rigby2022}, would be a major step forward in the study of clusters. First, the inclusion of IR wavelengths in the study of the stellar populations would constrain their properties (age and metallicity) better than using optical data alone \citep[e.g.,][]{Worthey1994}. Second, the IR explores the restframe optical/NIR wavelengths of more distant systems enabling a better characterization of the stellar mass distribution of these objects. 

The advent of JWST  marks a new era in the detection and study of the low surface brightness Universe. The exceptional IR sensitivity of JWST’s detectors holds great promise for the detection of extremely low surface brightness features. The aim of this work is to showcase, for the first time, the potential of JWST observations to study the diffuse light in clusters of galaxies.

Throughout this work, we adopt a standard cosmological model with the following parameters: $H_0=70$ km s$^{-1}$ Mpc$^{-1}$, $\Omega_m=0.3$ and $\Omega_\Lambda=0.7$. The redshift of this cluster is $z = 0.39$, corresponding to a spatial scale of 5.3 kpc/arcsec. All magnitudes in this paper are in the AB magnitude system.

\section{Data} \label{sec:data}

\begin{figure*}[ht!]
 \centering
   \includegraphics[width = 1.\textwidth]{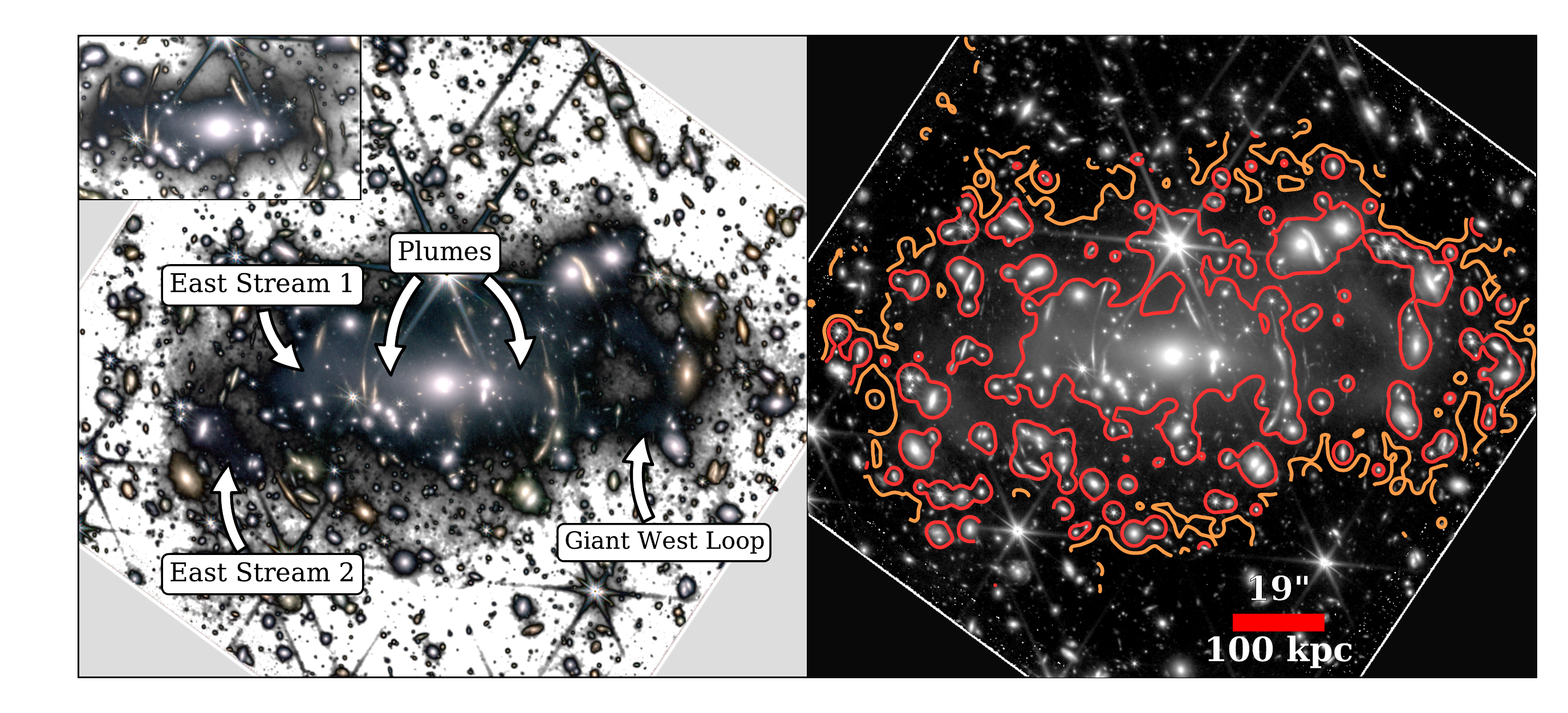}
   \caption{The cluster SMACS0723. The left panel shows an RGB composite of the long wavelength bands with a black and white background that is an average of the three filters. Some of the most prominent features in the ICL of the cluster are marked with  arrows. The inset on the top left is to better show the features in the center of the cluster. The right panel shows an average of the three long wavelength bands with two isocontours overplotted: 25 (red) and 27.5 (orange) mag/arcsec$^2$. North is up and East is left.  
\label{fig:smacs0723}}
\end{figure*}

The JWST Near Infrared Camera \citep[\nircam \footnote{\url{https://jwst-docs.stsci.edu/jwst-near-infrared-camera/nircam-instrumentation/nircam-detector-overview}},][]{Rieke2015} imaging is obtained at 0.6–2.3 $\mu$m (0.031\arcsec/pix, short wavelength channel) and 2.4-5.0 $\mu$m (0.063\arcsec/pix, long wavelength channel) \emph{simultaneously} over a 9.7 arcmin$^2$ field of view. \nircam{} uses a dichroic to observe in both channels in roughly the same field of view.
The camera consists of two modules (A and B) separated by a $\sim$44\arcsec gap. Each of the detectors contains $2048\times2048$ pixels. Only the inner $2040 \times 2040$ pixels are for science, and the outer 4-pixel wide border are reference pixels used for calibration. The short wavelength channel has 4 detectors, and therefore twice the resolution, and the long wavelength channel one, in each of the modules.

Deep observations of the cluster SMACS-J0723.3-7327 \citep[hereafter SMACS0723,][]{Repp2018} were taken with the \nircam{} filters F090W, F150W, F200W, F277W, F356W, and F444W. \nircam{} module B was centered on the cluster, and module A on an adjacent field. The cluster was observed on June 7th 2022 with several instruments aboard JWST as part of the observatory’s Early Release Observations \citep[ERO,][]{Pontoppidan2022}. The primary data set used for this work is based in the images obtained with \nircam, retrieved from the Mikulski Archive for Space Telescopes (MAST)\footnote{\url{https://outerspace.stsci.edu/display/MASTDATA/JWST+AWS+Bulk+Download+Scripts}}.

\subsection{Re-processing of the calibrated files}

The calibrated, co-added final  (\emph{Stage 3}) released images  present light gradients across the image, making them unsuitable to study the ICL of this cluster.

In this Section, we will explain our attempt at improving the data quality, re-reducing these images in a more suitable way for the purposes of this paper. In this  work, we concentrate our efforts on the long wavelength channel. The short wavelength channel needs a more elaborate re-reduction that is beyond the scope of this work. See Appendix \ref{app:shortw} for more details.

\subsubsection{Long wavelength channel}

The long wavelength channel consists of two detectors, one for each module. The pixel scale of this channel is $0.063\arcsec$/pixel.
The re-calibration of the long wavelength channel images were done in a two-step process. The light gradients in the final calibrated images (\emph{Stage 3}) are clearly an artefact. They are different from the ones observed in the short wavelength filters and not present in the RELICS HST images \citep{Coe2019}. The origin of these light gradients is not clear but they are observed both in the individual frames and the final co-add of all the long wavelength images.

To correct these gradients, first, we downloaded from MAST the calibrated (\emph{Stage 2}) individual frames of the F277W, F356W and F444W bands. For the first step of the process, we masked in each of the frames all galaxies, foreground and background objects and the central parts of the cluster. Then, we fitted a second degree 2D polynomial to the masked frame to model the gradient seen in the individual frames. The size of this plane is larger than the physical extent of the ICL in the images. This ensures that we are only correcting the gradient in the image while preserving the diffuse light of the cluster. In Appendix \ref{app:reduction}, we show an example of this process.

With the background of the individual images corrected, we created a preliminary co-add of the 9 exposure frames, per filter. This first co-add is then used to build a more accurate mask of the cluster, adding manually masked regions to cover the central parts of the image and minimize ICL contamination. This improved mask is then applied to the individual frames to obtain a more reliable model of the light gradient, for each of the frames. Once the frames are corrected from this gradient, we use SWarp \citep{Bertin2010} to create the final co-add. Finally, we computed and subtracted a constant background value from the final co-added images.

Note that we have used the original released images \citep{Pontoppidan2022}. It is known that there is a systematic magnitude offset due to the $20\%$ higher throughput of the filters in the long wavelength channel \citep{Rigby2022}. Therefore, we correct the photometry of the images using the offsets listed in \citet{Adams2022}. The offsets are: +0.223, +0.163 and +0.162 for the F277W, F356W and F444W bands, respectively. The zeropoints to convert to mag/arcsec$^2$ of each image are computed using the transformation in Appendix Sec. \ref{app:mags}.

The re-reduced images of the long wavelength channels are available through this link\footnote{ \url{https://www.dropbox.com/sh/zzb7fl1j0vl1vit/AABW1uz06Vn2XGCYAmOJIbega?dl=0}}. Although, we do not make use of them in this work, we have also made available the re-reduced images of the short wavelength channel. 

\subsection{Masking}

The study of the ICL in clusters of galaxies requires very careful masking of sources to reduce contamination that can affect the intrinsic properties of this light. In the case of deep images, this masking must be optimized for faint and small objects and for bright and large objects, background, foreground as well as belonging to the cluster. We use \sextractor's segmentation map \citep{Bertin1996} to build the masks for these images.

As a single setup for the detection and masking of both types of sources is unfeasible, we used a similar approach than in \citet{MT18, Montes2021}; two \sextractor{} runs: one to detect small sources (``hot'' mode) and another to detect larger, more extended objects (``cold'' mode). We use this approach on a median F277W + F356W + F444W image. 
In the case of the ``hot'' mode, we unsharp-masked the original image, to enhance the contrast of the smallest sources, particularly in the central parts of the cluster. To make the unsharp-masked image, we convolved the original image with a median box filter with a side of $5$ pixels and then we subtracted it from the original. This ``hot'' mask was further expanded 4 pixels. The brightest cluster galaxy (BCG) is left unmasked in both the ``hot'' and ``cold'' masks.  

The final mask was again visually inspected to manually mask any remaining light that was missed by the process described above. The spikes of the bright stars were also manually masked. In this final mask, all the objects are masked except for the BCG and ICL. 

\subsection{Surface brightness limits}

Our goal is to study the low surface brightness features in SMACS0723 down to the faintest surface brightness possible. For this reason, we need to know how deep our images are by estimating the surface brightness limits that they reach. To obtain these limits, we calculated the r.m.s per pixel of the final masked images and apply the formula given in \citet{Roman2019} to transform it to the typical $10\arcsec\times10\arcsec$ size. The $3\sigma$ surface brightness limits for the long wavelength channel are listed in Table \ref{table:sb}.

\begin{table*}[t]
\centering
\tabcolsep=0.2cm
\begin{tabular}{cccc}
\hline \hline
Filter & Channel &  Exp. Time  & Surface Brightness limits$^*$ \\
  &  & (seconds) & (mag/arcsec$^2$) \\
\hline
F277W  & Long & $7537.2$ &  31.28 \\
F356W  & Long & $7537.2$ &  31.32 \\
F444W  & Long & $7537.2$ &  31.10 \\
\hline
\end{tabular}
\caption{Summary of the \nircam{} observations used in this work. $^*$The surface brightness limits correspond to a sky fluctuation of 3$\sigma$ in an area of $10\times10$ arcsec$^2$.}
\label{table:sb}
\end{table*}

\section{The ICL of SMACS0723}
SMACS0723 is a cluster  at z = 0.39 discovered as part of the southern extension of the Massive Cluster Survey \citep{Repp2018}, with a mass of $M_h = 8.39^{+0.33}_{-0.34} \times10^{14} M_{\odot}$ \citep{Coe2019}. This cluster was selected for the JWST's ERO for its gravitational lensing potential, as shown by the prominent lensed arcs, and high latitude \citep[i.e., low Zodiacal emission,][]{Pontoppidan2022}.

Fig. \ref{fig:smacs0723} shows the final re-reduced images of SMACS0723. The left panel is an RGB color composite image of the F277W, F356W and F444W filters with an average of these three filters as a black and white background. The most notable features in the image are highlighted: two East Streams (1 and 2), the Giant West Loop and plumes East and West of the BCG. 
The East Stream 1 is located at 150 kpc from the BCG, while the East Stream 2 is at 243 kpc. The Giant West Loop covers a diameter of 100 kpc, from 150 kpc to 250 kpc from the center of the BCG. The width of the loop is quite homogeneous throughout the structure ($\sim30$ kpc). This suggests that it could be the remains of the tidal destruction of a galaxy of similar diameter. This will be discussed in the next sections.

The plumes seen in the center of the cluster span between 25 to 100 kpc from the BCG. They seem to be the result of the interaction of the BCG with a massive galaxy \citep{Lauer1988}. The inset in the left panel of Fig. \ref{fig:smacs0723} shows a zoom-in into the central part of the cluster to show the plumes more clearly.

The right panel of Fig. \ref{fig:smacs0723} shows the image average of the three filters with two isocontours: one representative of the galaxies of the cluster (red, 25 mag/arcsec$^2$) and one of the ICL at a distance of $\sim350$ kpc from the center of the BCG (orange, 27.5 mag/arcsec$^2$). 

\subsection{Radial surface brightness profiles of the BCG + ICL}

In this section we show the surface brightness radial profiles along the two semi-major axis of the cluster. This is done to highlight the asymmetries that are concentrated along this axis. We followed a similar approach as in \citet[][]{Montes2021}.

\begin{figure*}[ht!]
 \centering
   \includegraphics[width = 1\textwidth]{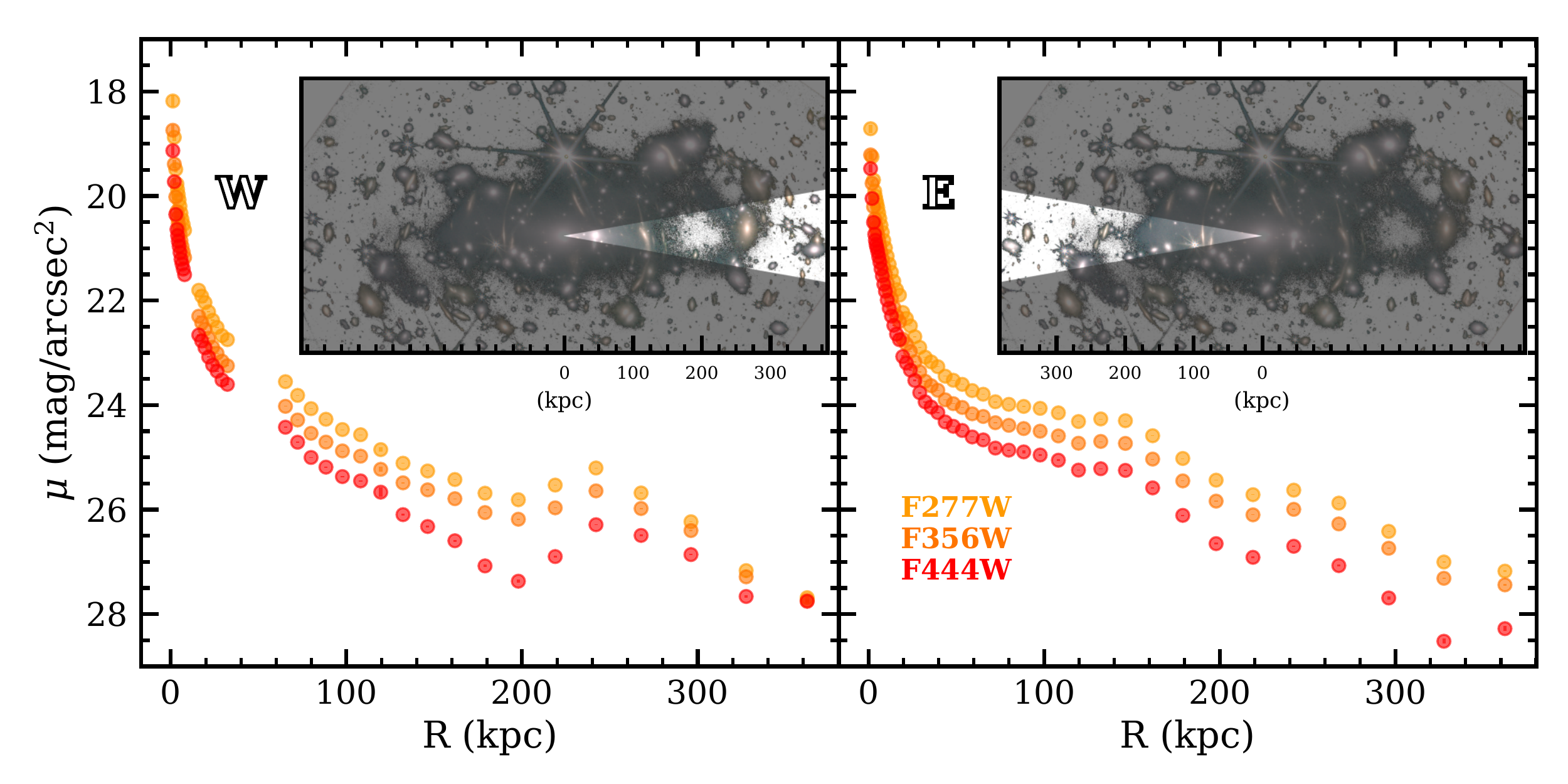}
   \caption{Surface brightness profiles as a function of radius for the BCG + ICL of SMACS0723, for the F277W (yellow), F356W (orange) and F444W (red) bands. Left panel shows the profiles towards the West while the right panel shows the profiles towards the East. The inset in both panels indicates the regions where the profiles were extracted. 
\label{fig:profiles}}
\end{figure*}

We masked everything except for a 20 deg-wide section towards the East and a 20-deg wide section to the West. We derived the radial profiles in each of the 3 bands in 51 circular logarithmic-spaced bins from 0 to 72 arcsec (or 380 kpc). The profiles for the East (right panel) and West (left panel) directions are shown in Fig. \ref{fig:profiles}. The insets in Fig. \ref{fig:profiles} show the region used to derive the profiles. The errors, smaller than the size of the markers, are computed as the quadratic sum of the r.m.s scatter of the signal and the r.m.s scatter of the background for each bin. The profiles are corrected for the extinction of the Galaxy \citep[E(B$-$V)=0.193,][]{Schlafly2011} using the \citet{Cardelli1989} extinction law. The gaps in the profiles correspond to masked regions in the images.

The surface brightness profiles are shown down to $\mu_{F444W}$ $\sim28$ mag/arcsec$^2$ ($\sim350$ kpc). This is around a factor of 20 above the limiting surface brightness of these images (Table \ref{table:sb}). We conservatively decided not to explore the ICL beyond $\mu_{F444W}$ $\sim28$ mag/arcsec$^2$ as the current background of the data is still affected by light gradients left during the re-reduction process. 
 
\subsection{Color and stellar mass density profiles}\label{sec:color}

\begin{figure*}[ht!]
 \centering
   \includegraphics[width = 0.9\textwidth]{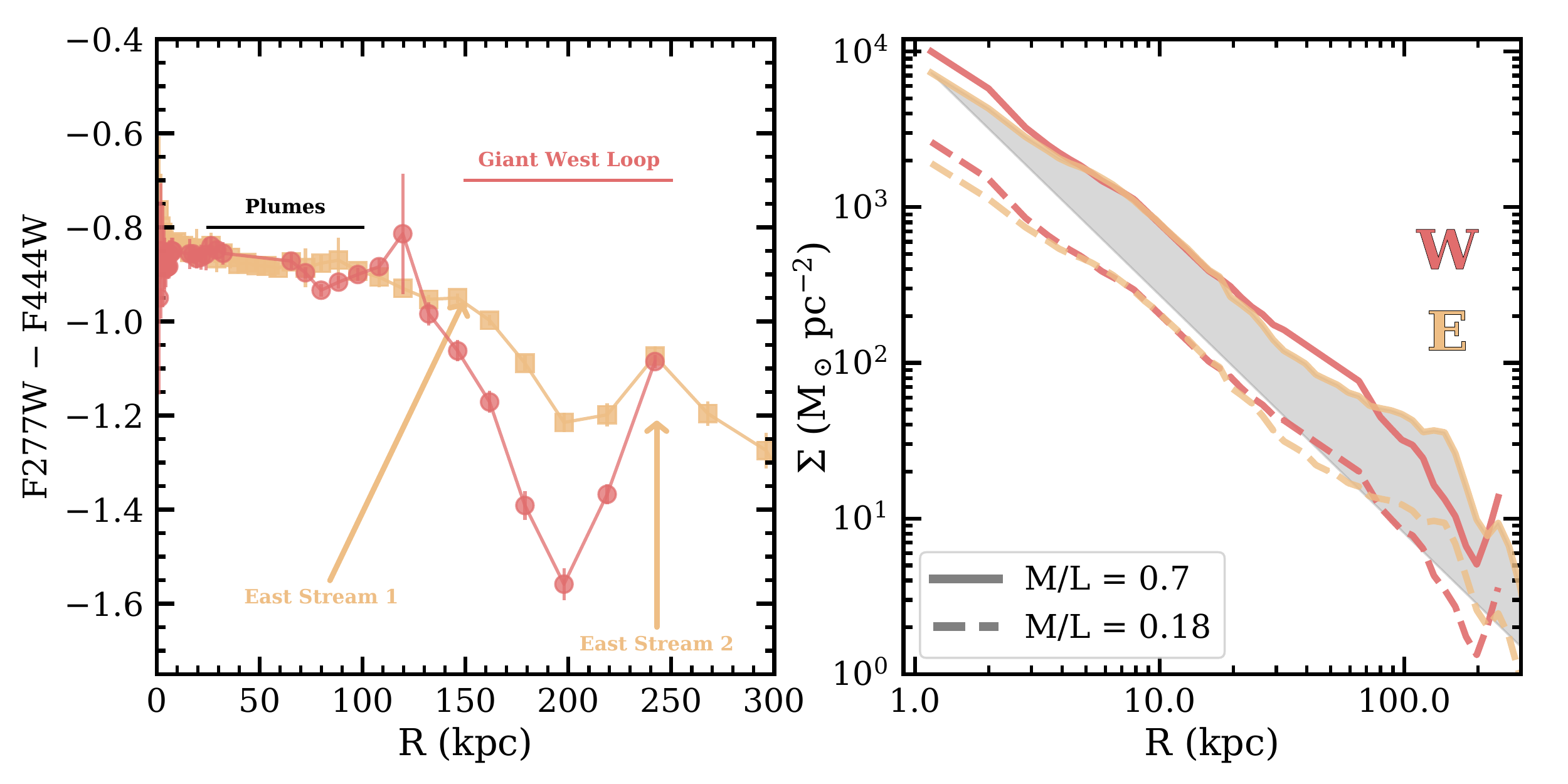}
   \caption{Left panel: the F277W$ - $F444W radial color profile of the BCG+ICL of SMACS0723. The features seen in the ICL are highlighted in the plot. Right panel: Stellar mass density radial profile of the BCG+ICL of SMACS0723. The solid line indicates the density profile derived using a M/L of 0.7 and the dashed line with a M/L of 0.18. The grey shaded area indicates the range of possible density profiles for this cluster. Note that the X-axis in this right-hand plot is in logarithmic scale.
\label{fig:color_dens}}
\end{figure*}

The left panel of Fig. \ref{fig:color_dens} shows the $F277W - F444W$ color profiles for the East (peach orange, squares) and West (pink, circles) profiles down to $300$ kpc from the center of the BCG. The errors are the combination of the errors from the radial profiles. As the outer parts of the images are more prone to systematics on the background subtraction, we conservatively show the radial color profiles down to a radial distance of $\sim300$ kpc. 

In the color profile plot, we also indicate the position of the features highlighted in Fig. \ref{fig:smacs0723}. For instance, the position of the Giant West Loop is evidenced by a wide ($\sim100$ kpc) dip of $\sim-0.6$ magnitudes in depth in the West profile. Instead, the East Streams appear as bumps in the East color profiles, indicating that the ICL is being enriched by the addition of redder stars. The plumes at each side of the BCG coincide with the almost constant color profile in the inner $\sim100$ kpc in both East and West color profiles; a signature of a major merger as the merger will lead to the flattening of the metallicity gradient of the BCG \citep[e.g.,][]{White1980}. 

The right panel in Fig. \ref{fig:color_dens} shows the stellar mass density profiles of SMACS0723 for the East (peach yellow) and West (pink) profiles. A restframe $3.6\mu$m filter would be the most representative tracer of the underlying stellar mass distribution \citep[e.g.,][]{Sheth2010}. Therefore, we used as a reference the F444W band as it is the closest to a $3.6\mu$m filter at the redshift of SMACS0723. To estimate the surface stellar mass density profile, we apply equation 1 in \citet{MT14}, to transform the F444W surface brightness profile to stellar mass density. We used the the mass-to-light ratios (M/L) predictions in the Spitzer IRAC $3.6\mu$m band from the E-MILES models \citep{Vazdekis2016}.

The BCG+ICL profiles of clusters at intermediate redshifts have negative gradients of both metallicity and age \citep[e.g.,][]{MT14, Morishita2017, MT18}. Lacking precise information about the age and metallicity of the BCG and ICL of this cluster, we took two M/L ratios: for an age of $7$ Gyr and [M/H] = $0.22$ ($0.70$, solid lines), and age $1.5$ Gyr and [M/H] = $-0.7$ ($0.18$, dashed lines) based on the values of age and metallicity of the Frontier Fields clusters for the centre and ICL from \citet{MT18}. By selecting these two extreme predictions for the M/L, we ensure that the most likely stellar mass density profile of the cluster is within the two lines in the right panel of Fig. \ref{fig:color_dens}.

Both stellar mass density profiles, East and West, follow a power-law profile. The profiles also present the different features highlighted in Fig. \ref{fig:smacs0723} like the Giant West Loop dip at $\sim200$ kpc and the bumps of the East Streams 1 ($150$ kpc) and 2 ($240$ kpc).  

\section{Discussion and conclusions}

The results presented in this work show the extraordinary potential of JWST to unravel the physical mechanisms behind the origin and evolution of the ICL. However, the current JWST NIRCAM data reduction is not yet low surface brightness-friendly. A tailored data reduction is a must in order to take the most advantage of the observations to study the faintest surface brightness features. 

\subsection{ICL formation in SMACS0723}

The asymmetry of the diffuse light in SMACS0723, along the East-West direction, means that the ICL of this cluster is still forming. The signature of this formation process can be seen in the prominent features that show up in both the surface brightness radial profiles in Fig. \ref{fig:profiles} and the radial color profile in the left panel of Fig. \ref{fig:color_dens}. In the following, we will discuss these features in detail.

\subsubsection{The Giant West Loop and the East Streams}

\begin{figure*}[ht!]
 \centering
   \includegraphics[width = 1\textwidth]{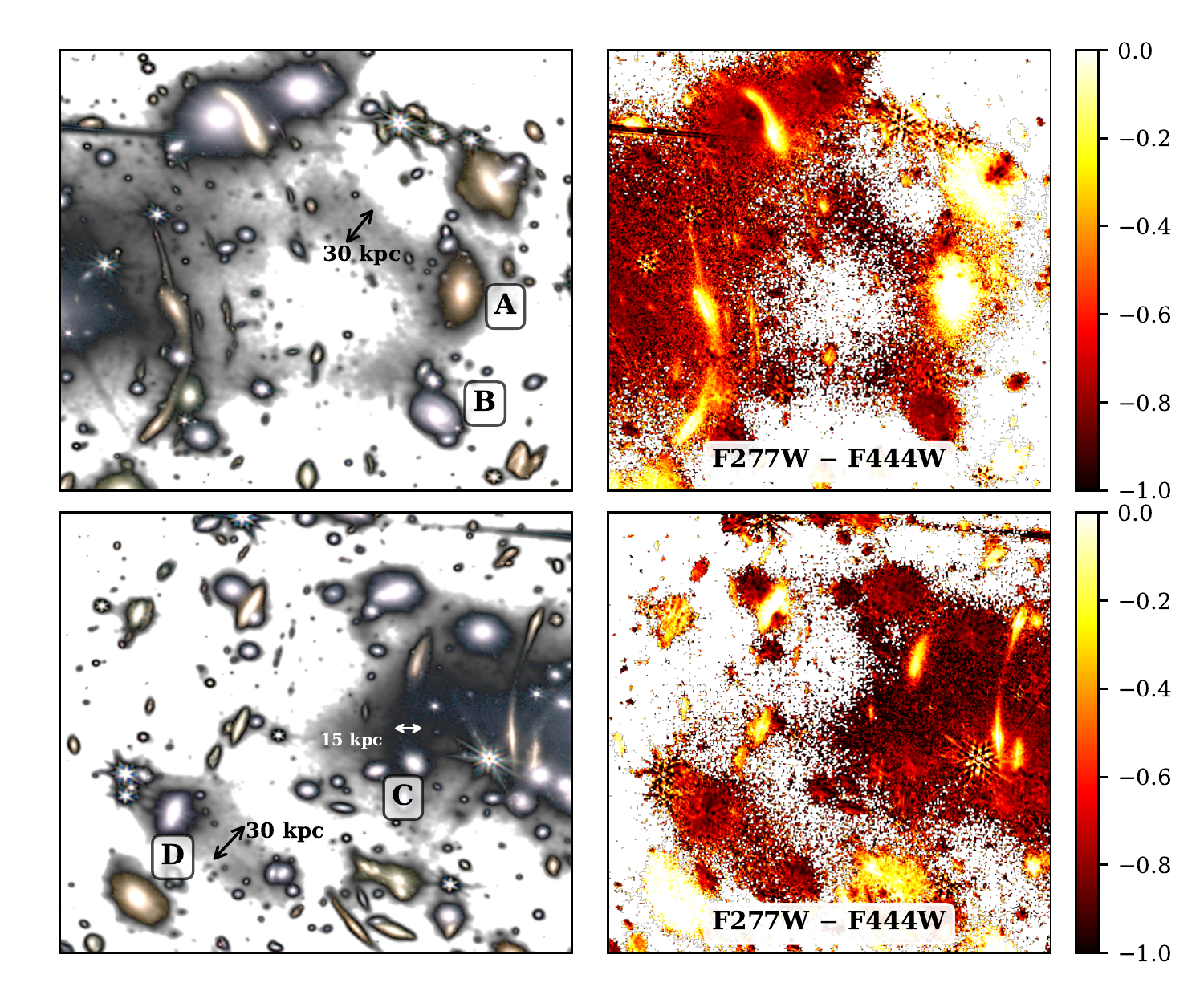}
   \caption{Zoom-ins into the different tidal features of SMACS0723. Left panels show the RGB composite images, while the right panels show the F277W$-$F444W color maps. The upper panels show the Giant West Loop and the lower panels the East Streams. The width of the different features is indicated with arrows. 
\label{fig:gwloop}}
\end{figure*}

Early observations showed that the ICL originates from the tidal stripping of satellite galaxies as they interact inside the cluster \citep[e.g.,][]{Gregg1998, Mihos2005}. This appears to be the origin of the features observed at larger radii, such as the East Streams and the Giant West Loop in Fig. \ref{fig:gwloop}. 

The circular shape of the Giant West Loop is composed by two parts. Its left side corresponds to the diffuse light from the center of the cluster rapidly decreasing. The right arc-like shape is likely the debris of a single stellar stream. The constant width of this feature ($\sim30$ kpc) and homogeneous color is consistent with being a single stellar stream. In Fig. \ref{fig:gwloop} upper left panel, we show the RGB color composite image of the region, labelling two of the galaxies that could potentially be associated with the loop.

The upper right panel shows the F277W$-$F444W color map of the same region. Although the North part of the loop seems to be associated with Galaxy A, this galaxy is redder than the stream and the galaxies of the cluster. In fact, it has a photometric redshift of 1.82 \citep{Coe2019}, placing it behind the cluster and not physically associated to the stream. Galaxy B, on the other hand, has colors that are compatible with that of the stream. The photometric redshift of this galaxy is 0.32 \citep{Coe2019}. However, the position of this galaxy is inconsistent with having created the entire Giant West Loop. Therefore, it remains an open question which galaxy has created this structure.

Both East Streams are shown in the lower panels of Fig. \ref{fig:gwloop}. The East Stream 1 is narrower (with a width of $\sim 15$ kpc) than the other features highlighted in this paper. In the color map, lower right panel, this feature does not show up as there is little contrast between the stream and the ICL background (see also Fig. \ref{fig:color_dens}). The color and width of this stream potentially associates it with Galaxy C. The East Stream 2 is wider ($\sim 30$ kpc), likely produced by the stripping of Galaxy D. This stream has a similar color than the ICL in the inner parts of the cluster, as seen in the color map.

The widths of both the Giant West Loop and the East Stream 2 are compatible with the size of Milky Way-like objects \citep[i.e. with a diameter of $\sim30$ kpc][]{Trujillo2020}. This is in agreement with the finding that the stars composing the ICL are mainly stripped from the outskirts of galaxies with stellar masses around $5\times10^{10}$ M$_\odot$ \citep{MT14, MT18, Morishita2017, DeMaio2018}.

\subsubsection{Plumes}

The plumes seen in the center of the cluster ($\sim25-100$ kpc, Fig. \ref{fig:smacs0723}) indicate that the BCG of this cluster is likely experiencing a merger \citep[see also][]{Duc2015, Martin2022}. At the same time, the measured color is flat in the inner $\sim100$ kpc in both East and West profile, indicating that the stellar populations in this region are well mixed. Flat gradients in color are indicative of stars expelled into the intracluster medium during a major merger \citep[$<$1:10, ][]{Krick2007, DeMaio2015, DeMaio2018}. The major merger is likely with the galaxy located 9 arcsec ($\sim47$ kpc) away to its West, as both galaxies are connected by a stellar bridge (zoom-in in Fig. \ref{fig:smacs0723}). Assuming similar stellar population properties between both galaxies, the merger ratio inferred is 1:3. 

\vspace{1cm}

The distinct features that compose the diffuse light of this cluster (plumes, streams, loop, etc) suggest different physical mechanisms creating the ICL (mergers and tidal disruption). This is telling us that the mechanisms that contribute to ICL formation do that differently depending on their distance to the centre of the cluster.

\subsection{The slope of the stellar mass density of the ICL}

The high degree of structure in the ICL of this cluster suggests that the diffuse extended component is being built now. Thanks to the dense environment of the cluster, in a few Gyrs these structures will smooth and become well-mixed \citep{Rudick2011}. Over time, the tidal stripping of more and more satellites will build the outer ICL \citep{DeMaio2018, DeMaio2020}.

To get an estimation of the slope of the stellar mass density profile, we have fitted a power-law to the BCG+ICL profile shown in Fig \ref{fig:color_dens}. As the centers of BCGs are usually old and high in metallicity, we use this fact as the starting point to fit the central regions. The slope and its errors encompass the possible solutions showed in Fig \ref{fig:color_dens}. The slope is $\alpha_{2D, E} = -1.48\pm0.13$, for the East profile, and $\alpha_{2D, W} = -1.46 \pm0.08$, for the West profile. Both values are in agreement within errors. That translates into a 3D slope of $\sim -2.47$ \citep[eq. 5 in][]{Stark1977}. The value of the 3D slope is in agreement with the values of the slopes in \citet{MT18} for the Frontier Fields clusters. This result implies that the stellar halo and the dark matter of SMACS0723 could have similar slopes \citep[the dark matter slopes range between $-2.6$ to $-2$,][]{Pillepich2014, Pillepich2018}, something expected for high-mass clusters \citep[$M_h = 8.39 ^{+0.33}_{-0.34} \times10^{14}M_{\odot}$,][]{Coe2019}.

\vspace{5mm}

In this work, we present the first analysis of the ICL of the ERO observations with JWST. We re-reduced the original JWST released images to minimize the inhomogeneities and make them more suitable for ICL studies. After our reprocessing, we measured that the ICL covers around half of the \nircam{} detector area. We derive the surface brightness radial profiles out to $380$ kpc from the center of the BCG, and the F277W$-$F444W color profiles to $300$ kpc. We found that:

\begin{itemize}
    \item The ICL of SMACS0723 presents a lot of substructure indicating that the processes that form the ICL in this cluster are caught in action.
    
    \item The color and morphological properties of the ICL in the inner $\sim100$ kpc are consistent with a major merger.
    
    \item The outer ICL ($>150$ kpc) is being formed by the tidal stripping of Milky Way-like galaxies.
    
    \item The slope of the stellar mass density profile of the BCG+ICL ($\alpha_{3D} = -2.47\pm0.13$) is nearly identical to the slope of the dark matter profile of clusters of this mass. This reinforces the idea of using the ICL as a potential dark matter tracer. Considering this, we encourage using the ICL to improve the gravitational lensing mass maps and, consequently, the characterization of the properties of the first galaxies.

\end{itemize}

\begin{acknowledgments}
This publication is part of the Project PCI2021-122072-2B, financed by MICIN/AEI/10.13039/501100011033, and the European Union “NextGenerationEU”/RTRP.
I.T. acknowledges support from the ACIISI, Consejer\'{i}a de Econom\'{i}a, Conocimiento y Empleo del Gobierno de Canarias and the European Regional Development Fund (ERDF) under grant with reference PROID2021010044 and from the State Research Agency (AEI-MCINN) of the Spanish Ministry of Science and Innovation under the grant PID2019-107427GB-C32 and  IAC project P/300624, financed by the Ministry of Science and Innovation, through the State Budget and by the Canary Islands Department of Economy, Knowledge and Employment, through the Regional Budget of the Autonomous Community.
The authors want to thank the JWST ERO team for making these extraordinary data available.
\end{acknowledgments}

%

\vspace{5mm}
\facilities{JWST (\nircam)}


\software{\texttt{astropy} \citep{Astropy2018},  
          Source Extractor \citep{Bertin1996}
          SWarp \citep{Bertin2010},
          \texttt{photutils} v0.7.2 \citep{Bradley2019},
          \texttt{pillow} \citep{pillow2020}, 
          \texttt{numpy} \citep{oliphant2006},
          \texttt{scipy} \citep{scipy2020}          
          }

\appendix

\section{Example of the data processing of the images from the long wavelength channel}\label{app:reduction}
\begin{figure*}[ht!]
 \centering
   \includegraphics[width = 1\textwidth]{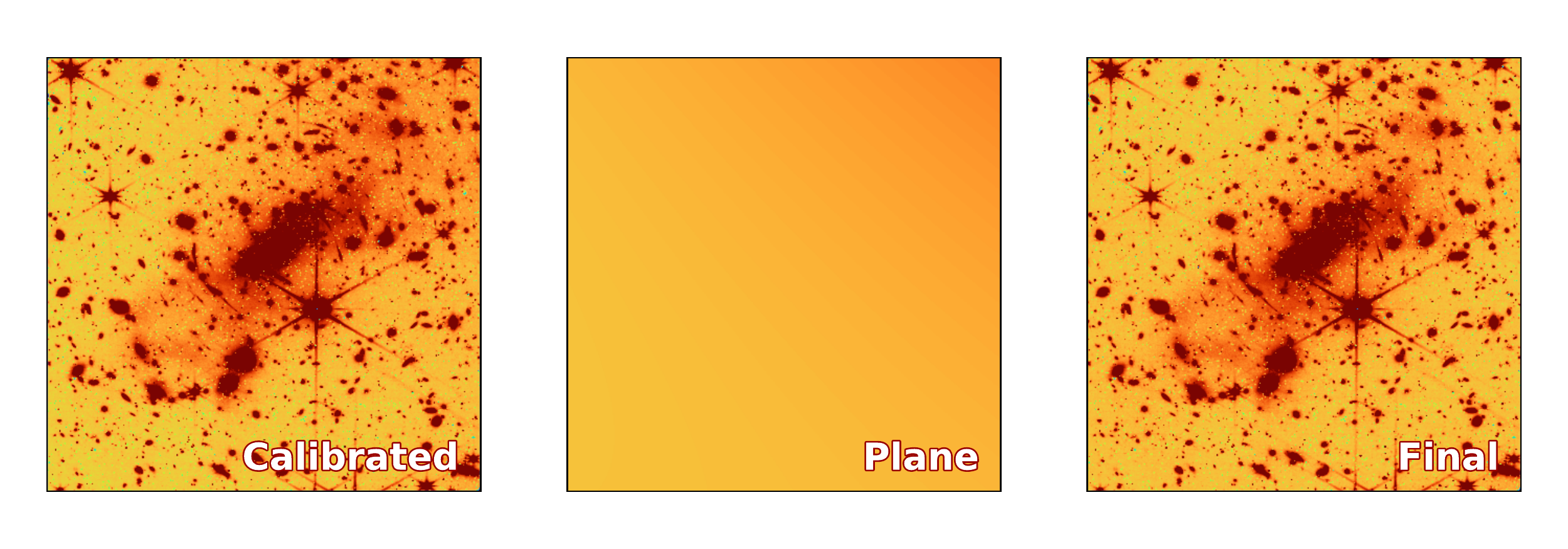}
   \caption{An example of a single frame  background correction. Left panel: Calibrated F277W frame (\texttt{jw02736001001\_02101\_00003\_nrcblong\_cal.fits}). A gradient across the image can be seen, towards the upper-right corner. Middle panel: 2nd degree polynomial plane fitted to the (masked) calibrated frame to correct the gradient. Right panel: Final frame after subtracting the plane.}\label{fig:plane}
\end{figure*}

In Sec. \ref{sec:data}, we detailed the data re-reduction that we performed in the JWST \nircam{} images. For the long wavelength channel, we fit a second degree 2D polynomial (a plane) to the calibrated images to eliminate the gradient seen across the image. The left panel in Fig. \ref{fig:plane} shows a \texttt{Stage 2} calibrated frame. The middle panel shows the plane fitted to the masked original calibrated frame. The right panel shows our final result, without the gradient. 

\begin{figure*}[ht!]
 \centering
   \includegraphics[width = 1\textwidth]{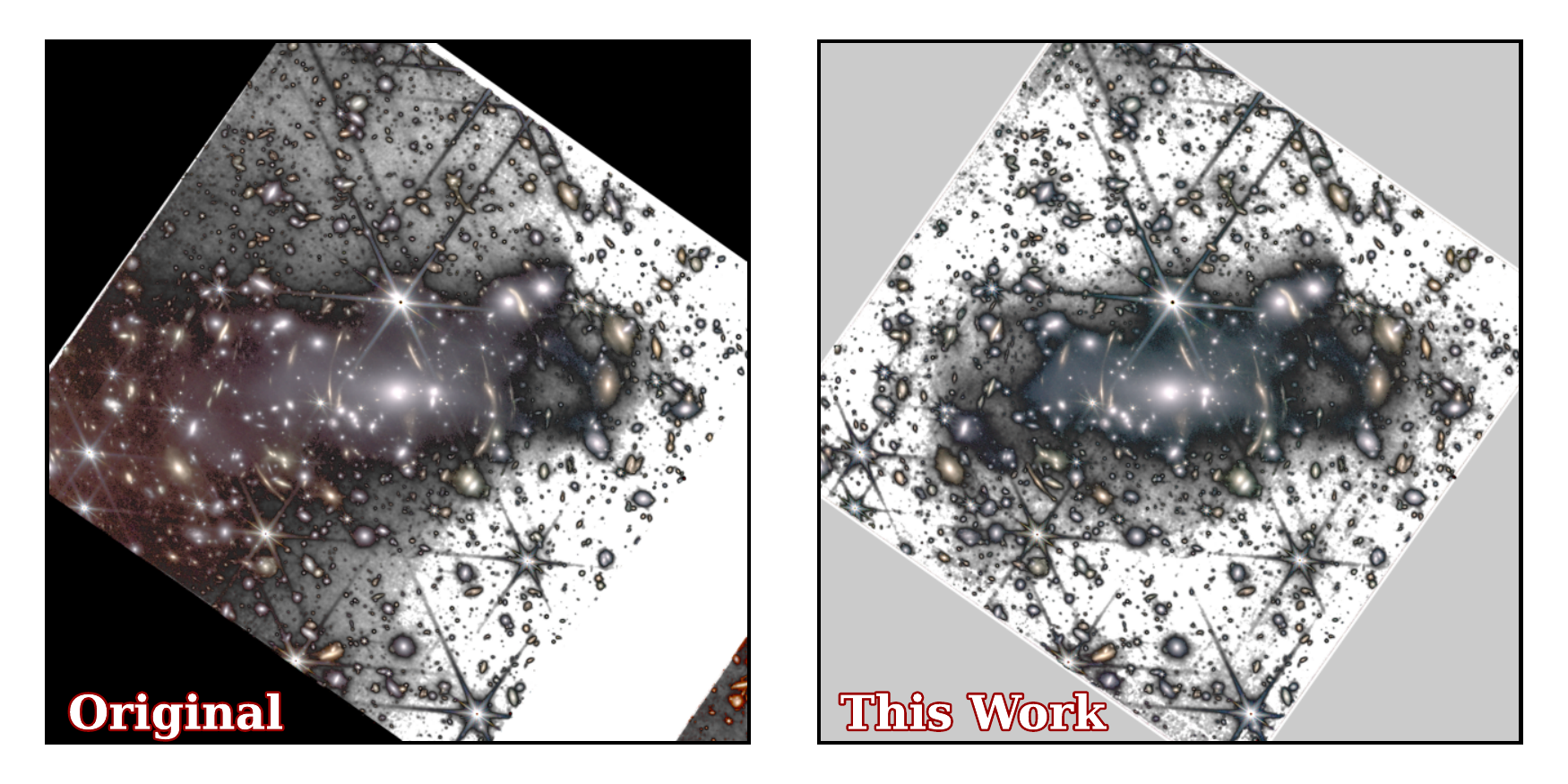}
   \caption{Left panel: RGB color composite image using the original calibrated co-adds of the long wavelength channels. Right panel: RGB color composite image using the re-reduced co-add images created in this work. Note the huge improvement of correcting the strong light gradient towards the East side on the NIRCam camera. }\label{fig:complong}
\end{figure*}

Fig. \ref{fig:complong} shows the improvement of the data reduction performed here. Both panels show RGB color composite images using the long wavelength bands and with an average of the three filters as the black and white background. The original images (in the left) are those using the JWST current reduction pipeline whereas our re-reduction is shown in the right panel.

\section{Short wavelength channel re-reduction}\label{app:shortw}

The short wavelength channel consists of a mosaic of 8 detectors (4 for each module), manufactured with a 2.5 $\mu$m wavelength cutoff. In each module, the 4 short wavelength detectors are arranged in a $2 \times 2$ array. The pixel scale of this channel is $0.031\arcsec$/pixel.

Contrary to the case of the long wavelength channel, the detectors in the short wavelength channel are smaller than the extent of the diffuse light of the cluster. For this reason, fitting a plane to each one of the detectors is risky, more so when the detectors do not show any common gradient. In this case, we opted for a more conservative approach: to correct for the readout pattern per detector \citep[see][]{Pontoppidan2022, Merlin2022} and scale the 4 detectors.

\begin{figure*}[ht!]
 \centering
   \includegraphics[width = 1\textwidth]{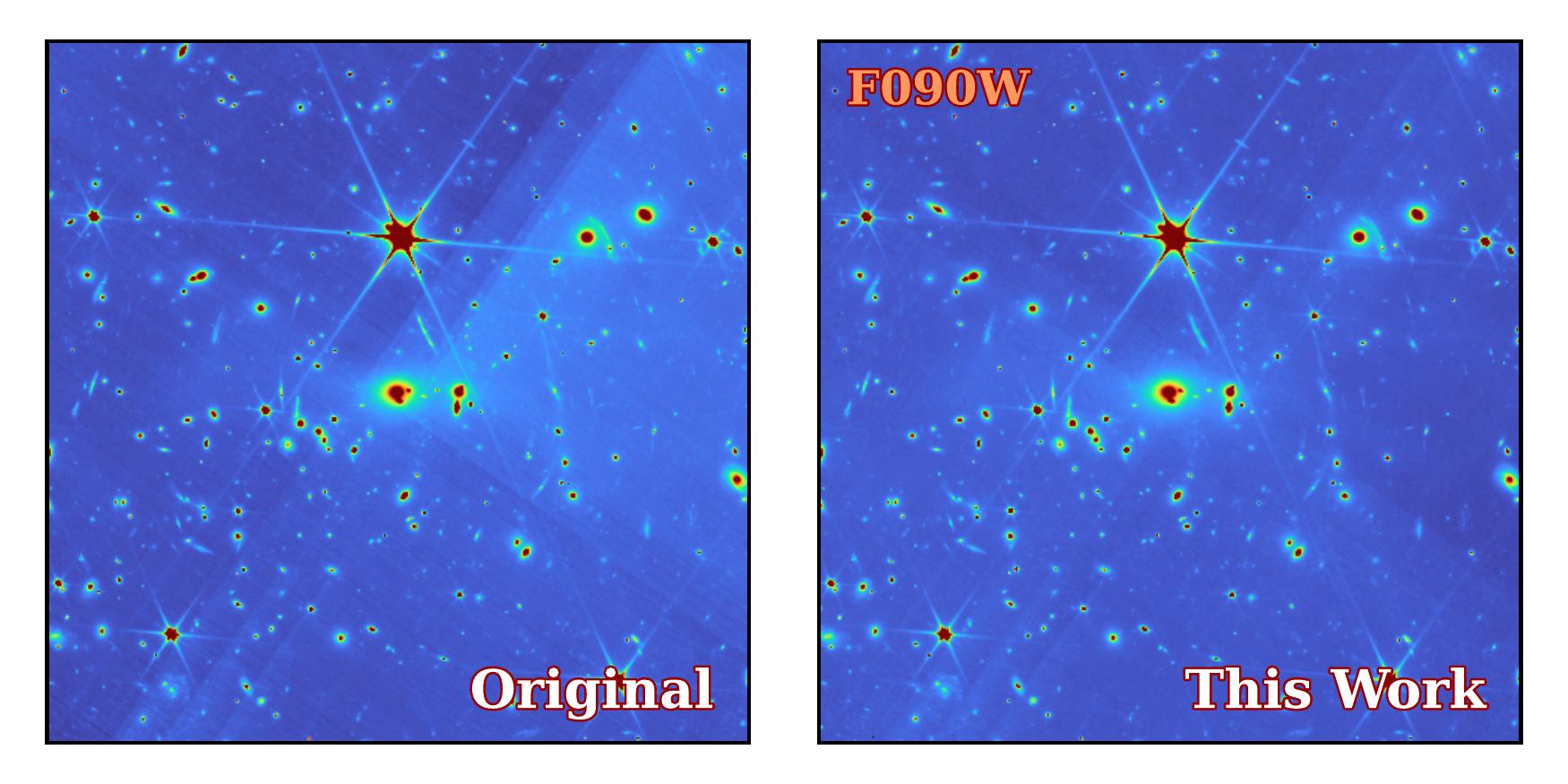}
   \caption{Left panel: Original calibrated F090W co-add. Right panel: Final co-add after the data re-reduction. \label{fig:f090}}
\end{figure*}

One of the characteristics of the \nircam{} detectors is the presence of low-level electronic noise manifesting as linear stripes \citep[`1/f noise', see also][]{Pontoppidan2022}. The released ERO images are already corrected from this noise using a simple correction. However, residuals are left, clearly visible in the short wavelength channel images (left panel Fig. \ref{fig:f090}). 

The first step was to remove this remaining pattern by subtracting the $3\sigma$-clipped median value for each row, horizontally, in the calibrated individual frames after masking out all the objects. Once this correction was applied, we derived an improved mask from the corrected images and compute and subtract the median value for each row. 

Once the images per detector were corrected from the remaining `1/f noise', we built the mosaics. The left panel in Fig. \ref{fig:f090} shows that in the  final calibrated mosaics the different detectors were not properly scaled. Therefore, we have also scaled the mosaics by multiplying each detector by a factor, taking detector one (\texttt{nrcb1}) as reference. This factor is computed as the $3\sigma$-clipped median, for each frame, in a ring of radius $\sim47\arcsec$ and width of $6\arcsec$ from the center of the frame, after masking all sources. After scaling the different detectors, the frames were assembled using \texttt{SWarp}. The final output is created as the median of the 9 mosaic frames. 

An example of the data reduction for the F090W band is shown in Fig. \ref{fig:f090}. For comparison, we show the original image in the left panel, and our data reduction in the right panel. Despite the clear improvement, the final images of the short wavelength channel still show significant residuals that prevent us to use them to study the ICL. These residuals seem to be caused by inhomogeneities in the 4 different adjacent sectors ($512 \times 2048$ pixels) that form each of the detectors\footnote{\url{https://www.stsci.edu/files/live/sites/www/files/home/jwst/documentation/technical-documents/_documents/JWST-STScI-001721.pdf}}.

\section{From MJy/sr to surface brightness in JWST data}\label{app:mags}

In this work, all the units of the photometry, surface brightness and color profiles, are given in mag/arcsec$^2$ (in the AB system). The explicit equation to transform from MJy/sr (the units of the JWST images) to surface brightness in the AB system is:

\begin{equation}
    \mu_{AB} = -2.5\times \log_{10} (\mathrm{Counts}\, [\mathrm{MJy/sr}] \times \frac{10^6} {4.25\times10^{10}}) + 8.906
\end{equation}

Where 4.25$\times$10$^{10}$ corresponds to number of arcsec$^2$ in 1 sr. The factor 10$^6$ stands for the transformation from MJy to Jy. Finally, the 8.906 value corresponds to the AB system referred to 3631 Jy as the reference unit \citep{Oke1983}.


\bibliography{icl}{}
\bibliographystyle{aasjournal}



\end{document}